\UseRawInputEncoding 
\documentclass[%
preprint,
 amsmath,amssymb,
 aps,
]{revtex4-2}

\usepackage{graphicx}
\usepackage{dcolumn}
\usepackage{bm}

\begin{document}


\title{Quantum Molecular Charge-Transfer Model for Multi-step Auger-Meitner Decay Cascade Dynamics}

\author{A. E. A. Fouda}
\affiliation{Department of Physics, The University of Chicago, Chicago, IL 60637, USA}
\affiliation{Chemical Sciences and Engineering Division, Argonne National Laboratory, 9700 S. Cass Avenue, Lemont, IL 60439, USA}
\email{adamfouda@uchicago.edu}
\author{S. H. Southworth}  
\affiliation{Chemical Sciences and Engineering Division, Argonne National Laboratory, 9700 S. Cass Avenue, Lemont, IL 60439, USA}
\author{P. J. Ho}  
\email{pho@anl.gov}
\affiliation{Chemical Sciences and Engineering Division, Argonne National Laboratory, 9700 S. Cass Avenue, Lemont, IL 60439, USA}

\begin{abstract}
The fragmentation of molecular cations following inner-shell decay processes in molecules containing heavy elements underpins the x-ray damage effects observed in x-ray scattering measurements of biological and chemical materials, as well as in medical applications involving Auger-electron-emitting radionuclides. Traditionally, these processes are modeled using simulations that describe the electronic structure at an atomic level, thereby omitting molecular bonding effects. This work addresses the gap by introducing a novel approach that couples a decay spawning dynamics algorithm with \textit{ab initio} molecular dynamics simulations to characterize ultrafast dynamics on the potential energy surfaces.  We apply our method to a model decay cascade following K-shell ionization of IBr and subsequent K$_{\beta}$ fluorescence decay. We examine two competing channels that undergo two decay steps, resulting in ion pairs with a total +3 charge state. This approach provides a continuous description of the electron transfer dynamics occurring during the multi-step decay cascade and molecular fragmentation, revealing the combined inner-shell decay and charge transfer timescale to be approximately 75 fs. Our computed kinetic energies of ion fragments show good agreement with experimental data.
\end{abstract}

\maketitle

\section{Introduction}\label{sec:intro}

Observing ultrafast molecular dynamics often involves the detection of molecular cation fragments and electrons following inner-shell decay processes, however little is known about the explicit nature of the molecular breakup dynamics\cite{PhysRevX.6.021035}. The structural rearrangements following the decay of core-hole states have far-reaching implications. They are the fundamental process behind the radiation damage effects caused by structural distortions occurring during x-ray scattering measurements of metallo-protein and clean energy materials\cite{Garman:me0638, doi:10.1080/09553008614551071, carugo2005x}. Also, new highly targeted cancer therapies are being developed that exploit Auger-Meitner decay cascades initiated by radionucleotides bonded to small carrier molecules\cite{Ku2019,PIROVANO202150}. Furthermore, Coulomb explosion imaging that uses intense x-ray pulses to create highly charged fragments have been employed to record molecule structures \cite{Rudenko2017, Boll2022, Li-PRL-2022}. Therefore, there is demand for the development of simulations that can explicitly track both the ejection of electrons and the associated structural and fragmentation dynamics of x-ray excited molecules. 

Synchrotron experiments employing x-ray/ion coincidence spectroscopy can discern the charge states and kinetic energies of diverse decay channels subsequent to x-ray-induced molecular breakdown, as noted in \cite{10.1063/5.0145215}. These experiments serve as reliable benchmarks for emerging quantum molecular dynamics methods. We previously employed Monte Carlo/Molecular Dynamics (MC/MD) simulations combined with the classical over-the-barrier (COB) model \cite{Ho_2017,PhysRevLett.113.073001,10.1063/5.0145215} to study the fragmentation dynamics in IBr molecule induced by x-rays. This approach enables tracking of inner-shell cascades and incorporates electron redistribution and nuclear motion across the ground, core-hole, and valence states for all feasible charge configurations. However, it employs the atomic Hartree-Fock-Slater (HFS) method to describe the electronic structure, and the electron transfer via the COB model promptly fills vacancies in neighboring atoms when the electron binding energy exceeds the Coulomb barrier. Consequently, molecular bonding effects, as well as the quantum effects of electron transfer, are absent from these simulations.

Although there has been recent progress in the computation and interpretation of molecular Auger-Meitner decay spectra\cite{10.1063/5.0036977, grell2020multi, cabraljctc2022, D3CP01746J, D4CP00053F, doi:10.1021/acs.jpclett.3c03611}, there have been few developments in quantum molecular dynamics methods for this decay process. Coupling nuclear dynamics to Auger-Meitner decay cascades poses challenges because of the exponential increase in possible final states and potential energy surfaces with system size. Furthermore, the decay of core-hole states in heavy elements, such as bromine and iodine, involves multiple cascade steps, further increasing the complexity of decay dynamics that couples a multitude of potential energy surfaces. Picon introduced a theoretical framework using numerical time-dependant Shr\"{o}digner equation simulations\cite{PhysRevA.95.023401}, applied to in a recent x-ray pump/probe study\cite{Vassholz2021}. Other studies have focused on vibronically resolving Auger-Meitner decay and coupling it with conformer sampling from dynamics simulation\cite{Takahashi_2009,Takahashi_2015,10.1063/1.2166234,D1CP05662J}. To our knowledge, no time-resolved studies have yet coupled nuclear dynamics with a multi-step, Auger-Meitner decay dynamics in a heavy-element-containing molecule.


In this work, we provide a quantum molecular description for modeling charge transfer and decay dynamics during the structural damage of molecules following x-ray interactions. To capture the decay dynamics, our approach utilizes a time-dependent ensemble of quantum amplitude trajectories that transfers population by spawning trajectories across potential energy surfaces with increasing charge state.  Our approach is inspired by the \textit{ab-intio} multiple spawning (AIMS) method, which uses a time-dependant set of trajectory basis functions to model the nonadiabatic dynamics associated with the population transfer near conical intersections\cite{doi:10.1021/jp994174i,doi:https://doi.org/10.1002/0471264318.ch7,doi:10.1021/acs.chemrev.7b00423,fedorov2020pyspawn}.  To describe the electron reorganization and redistribution in the valence shells and the subsequent charge separation dynamics, we employ \textit{ab-inito} molecular dynamics (AIMD)\cite{iftimie2005ab} simulation coupled to a novel decay spawning dynamics algorithm. To validate our approach, we use a simplified model to investigate the decay cascade of the IBr molecule, beginning with a K-shell ionization followed by a K$_{\beta}$ fluorescence decay in Br atom.  This cascade dynamics has been measured experimentally using an x-ray/ion coincidence.  Our calculated results show good agreement with the measured kinetic energy released of the ions and branching ratio of fragmentation channels.

Our paper is organized as follows.  In Section \ref{sec:model}, we will describe the decay cascade model for the molecular dynamics of IBr. In Section \ref{sec:eqs} and \ref{sec:comp}, we present the theoretical treatment and computational details respectively and in Section \ref{sec:algo}. Finally, in Section \ref{sec:results}, we show the calculated results, followed by a summary and outlook.

\section{Theoretical Method}\label{sec:theory}




In this section, we present the theoretical framework, electronic structure theory and computational model for treating x-ray induced electron and molecular dynamics. To validate the algorithm and workflow of our model, we compute observables for direct comparison with the experimental data of the IBr molecule.


The experiment was conducted at beamline 7-ID at the Advanced Photon Source, utilizing x-ray/ion coincidence spectroscopy to measure the charge states and energies of I$^{q+}$ and Br$^{q'+}$ atomic ions following 1s ionization at the I and Br \textit{K}-edges of IBr. Detailed experimental procedures are available in our previous work \cite{10.1063/5.0145215}, where we presented ion data in coincidence with K${\alpha}$ emission from both I and Br atoms, along with the mechanisms for ion production derived from the MC/MD method. In the current study, we investigate the decay cascade of the IBr molecule followed by K${\beta}$ fluorescence decay in the Br atom. We focus on the dynamics leading to the formation of $I^{2+}/Br^{+}$ and $I^{+}/Br^{2+}$. The experimental data were recorded for both $^{79}$BrI and $^{81}$BrI and similar results were obtained for both isotopes. For our calculations, we uses the $^{79}$BrI isotope, for which the the experimental relative yield of $I^{2+}/Br^{+}$ to $I^{+}/Br^{2+}$ is 1.5 and measured kinetic energy releases were 12.2 and 11.9 eV, respectively.


\subsection{Decay Cascade Model}\label{sec:model}

\begin{figure}[ht!]
 \centering
 \includegraphics[width=10cm]{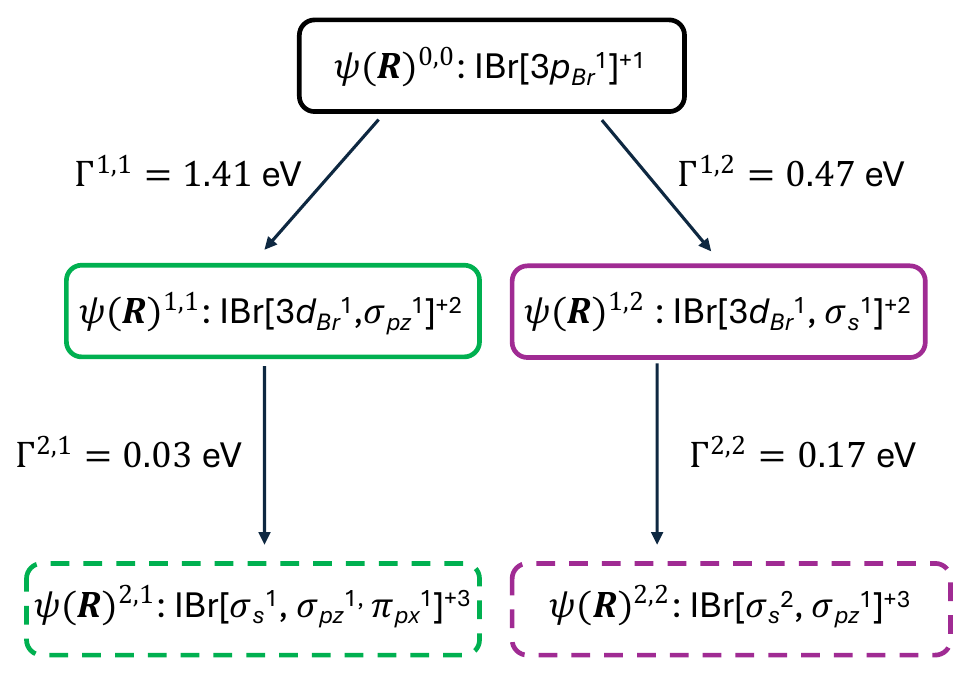}
 \caption{Schematic of the model decay cascade used in this work. The orbital label superscript indicates the hole occupation.}
 \label{fgr:model}
\end{figure}

We use a simplified model to characterize the molecular dynamics.  Using MC/MD simulations, we identified two dominant and competing decay channels and molecular states of IBr, starting from a core-excited state with a 3p vacancy in Br atom. Both channels lead to a molecular ion with a total charge state of 3+ via two Auger decays.  We index the states based on which step ($I$) and channel ($J$) they occur in and label the states as $|\psi(\textbf{R})^{I,J}\rangle$. The initial Br 3$p$ hole cation state is considered to be the zeroth step of the zeroth channel and is labelled as the $|\psi(\textbf{R})^{0,0}\rangle$ state. In the first decay step, the $|\psi(\textbf{R})^{0,0}\rangle$ state decays to the $|\psi(\textbf{R})^{1,1}\rangle$ and $|\psi(\textbf{R})^{1,2}\rangle$ states in channels 1 and 2 respectively, as shown in Fig. \ref{fgr:model}. $|\psi(\textbf{R})^{1,1}\rangle$ has a vacancy in both the 3$d$ orbital of Br atom and $\sigma_{pz}$ valence orbital. On other hand, $|\psi(\textbf{R})^{1,2}\rangle$ has a vacancy in both the 3$d$ and $\sigma_{s}$ orbital. In the second decay step, $|\psi(\textbf{R})^{1,1}\rangle$ and $|\psi(\textbf{R})^{1,2}\rangle$ decays to $|\psi(\textbf{R})^{2,1}\rangle$ and $|\psi(\textbf{R})^{2,2}\rangle$  respectively. $|\psi(\textbf{R})^{2,1}\rangle$ has a vacancy in $\sigma_{s}$ orbital and two holes in the valence $\sigma_{pz}$ and $\pi_{px}$ orbitals, whereas $|\psi(\textbf{R})^{2,2}\rangle$ has two holes in $\sigma_{s}$ orbital and one hole in the valence $\sigma_{pz}$ orbital. The associated Auger-Meitner decay rates ($\Gamma$)shown Fig. \ref{fgr:model} are obtained using HFS simulations, which are discussed in detail elsewhere\cite{Ho_2017,10.1063/5.0145215}. We point out the possible fluorescence decays, which take at picoseconds or longer, after step 2 are not included in the current simplified model.


\subsection{Coupled AIMD and Decay Spawning Dynamics}\label{sec:eqs}

This subsection describes the workflow of our coupled decay spawning dynamics and AIMD method and its capability for calculating experimental observables.
For each of the states in the decay channels, a single non-interacting potential energy surface is used to approximate all possible electronic configurations. 
In this study, we precalculate all the potential energy surfaces and use them to model the Auger-Meitner decays. Therefore, preliminary AIMD simulations were performed to produce the  non-interacting potential energy surfaces of selected non-ground state cation configurations of the electronic structure.
We will discuss the electronic structure calculations in the computational details section.  AIMD is a mixed classical-quantum approach which enables the characterization of potential energy surfaces ``on-the-fly", by classically propagating the nuclei in time with forces determined by an approximate solution to the time-independent electronic Schr\"{o}digner equation,
\begin{equation}\label{eq:aimd}
M_{Nuc}\textbf{\"{R}}_{Nuc}(t) = -\nabla_{Nuc}[\min_{\psi(\textbf{R})}\{\langle\psi(\textbf{R})|\hat{H}_{e}|\psi(\textbf{R})\rangle\} + V_{NN}(\textbf{R})].
\end{equation}
$\textbf{R}$ denotes the nuclear coordinates, $\hat{H}_{e}$ and $\psi(\textbf{R})$ are the electronic Hamiltonian and electronic state wave-functions, and $\min\limits_{\psi(\textbf{R})}$ indicates that the electronic structure undergoes self-consistent field optimization at the new nuclear coordinates produced at each time-step. The term $V_{NN}(\textbf{R})$ represents the nuclear-nuclear repulsion component of the potential energy. Evaluating this equation allows for the propagation of nuclear motion classically by numerically solving the Velocity Verlet algorithm \cite{10.1063/1.442716}. This step yields both the time-dependent nuclear kinetic energies and the atomic charges and predicts the final kinetic and charge states of the ions, which are the relevant experimental observables in x-ray/ion-coincidence experiments. 

Characterizing the non-interacting potential energy surfaces using AIMD facilitates the formation of a total electronic wavefunction for all states involved in the decay cascade model. This wavefunction is parametrically dependent on the nuclear coordinates. Within the Markov approximation, we represent this wavefunction by separating the time-dependent coefficients from their stationary state oscillation in time, as shown below,
\begin{equation}\label{eq:wf1}
|\Psi(\textbf{R},t)\rangle  = C^{0,0}(t)|\psi(\textbf{R})^{0,0}\rangle + \sum^{N_{I}}_{I}\sum^{N_{J}}_{J}C^{I,J}(t)|\psi(\textbf{R})^{I,J}\rangle 
\end{equation}
This wavefunction anzatz sums over the number of states in the steps ($I$) and channels ($J$) of the decay cascade. The indexes on the electronic state wavefunction indicate the decay steps and channels.  For our IBr study showed in Fig. \ref{fgr:model}, both $N_{I}$ and $N_{J}$ equal 2, and $|\psi^{0,0}(\textbf{R})\rangle$ is the initial IBr[$3p_{Br}^{1}$]$^{+1}$ state. 

To model the Auger-decay induced population transfer between states, our method utilizes a time-evolving basis of amplitude trajectories that propagate across the predefined AIMD potential energy surfaces. We adopt the spawning nomenclature from AIMS by stating that the $|\psi^{0,0}(\textbf{R})\rangle$ state is the parent state with its parent amplitude trajectory $C^{0,0}(t)$. The child amplitude trajectories ($c_{i}^{J,J}(t)$) for states with $I = 1$ will be spawned from $C^{0,0}(t)|\psi^{0,0}(\textbf{R})\rangle$, and all states with $I > 1$ will have child amplitude trajectories spawned from the $I-1$ states of the corresponding channel. Therefore, there will be only a single amplitude trajectory for $|\psi^{0,0}(\textbf{R})\rangle$, described by $C^{0,0}(t)$. The quantum amplitude trajectories of the states with $I > 1$ will be expressed as linear combinations of the child trajectories and are given as follows:
\begin{equation}\label{eq:child}
C^{I,J}(t)\psi(\textbf{R})^{I,J} = \sum^{N^{I}(t)}_{i}c_{i}^{I,J}(t)|\phi_{i}(\textbf{R})^{I,J}\rangle
\end{equation}
$\phi_{i}(\textbf{R})^{I,J}$ indicates the child trajectory states for each electronic state.  For each value of $I$, there will be the same number of trajectories ($N^{I}(t)$) in all channels throughout the simulation. 

Individual quantum amplitudes, $C^{0,0}(t)$ and $c_{i}^{I,J}(t)$, evolve across the potential energy surfaces predefined by the AIMD simulations and are
determined by the following equations of motion for the Auger-Meitner decay,
\begin{equation}\label{eq:eom1}
i\dot{C}^{0,0}(t) = \Bigl(E(\textbf{R},t)^{0,0}-\frac{i\Gamma^{1,\bar{J}}}{2}\Bigl) c^{0,0}(t) .
\end{equation}
and 
\begin{equation}\label{eq:eom2}
i\dot{c}_{i}^{I,J}(t) = \Bigl(E(\textbf{R},t)^{I,J}_{i}-\frac{i\Gamma^{I+1,\bar{J}}}{2}\Bigl) c_{i}^{I,J}(t) .
\end{equation}
$E(\textbf{R},t)^{I,J}_{i}$ is the energy of the trajectory state at its point on the potential energy surface. In the current model, coupling between potential energy surfaces is ignored and the population transfer between the incremental charge states is governed by the depletion of population by $\Gamma^{I,\bar{J}}$. Here $\Gamma^{I,\bar{J}}$ (note $\bar{J}$) indicates the total decay rate which is the sum of all possible decay channels for a state ($\Gamma^{I,\bar{J}} = \sum_{J} \Gamma^{I,J}$). In the case that nonlocal decays are important, $\Gamma^{I,\bar{J}}$ will have a $\textbf{R}$ dependence.

The population transfer from state $(I,J)$ to state $(I+1,J)$ is determined by the following loss function\cite{doi:10.1080/00268976.2022.2133749},  
\begin{equation}\label{eq:loss}
b_{i}^{I,J}(\tau_{n+1}) = \Gamma^{I+1,J}\int_{\tau_{n}}^{\tau_{n+1}}dt p_{i}^{I,J}(t) ,
\end{equation}
where the trajectory state population ($p_{i}^{I,J}(t)$) is defined by,
\begin{equation}\label{eq:trajpop}
p_{i}^{I,J}(t) = |c_{i}^{I,J}(t)\Big|^{2}.
\end{equation}
$\tau_{n}$ is the spawning recurrence parameter and the integral between $\tau_{n+1}$ and $\tau_{n}$ is the population loss occurring between spawning events. The population loss $b_{i}^{I,J}(t)$ will become the population gained on state $|\psi(\textbf{R})^{I+1,J}\rangle$, by spawning a child trajectory function that inherits the properties of its parent trajectory. In particular, the initial molecular structure of the child trajectory will have the same or similar structure as the parent trajectory at the time of the spawning event. This is based on the assumption that the Auger-Meitner decay process continually depletes the population adiabatically and incur little or no structural changes during the population transfer.


$\tau_{n}$ is a user defined parameter that controls the resolution of the wave-packet evolution. 
If the user was only interested in the final state populations and properties, then $\tau_{n}$ would only occur at the final time-step of the simulation. However, in order to capture the charge transfer time of the coupled Auger-Meitner decay and fragmentation dynamics, the selection of $\tau_n$ should capture Auger-Meitner decay rates so that the transfer of population between the states is visualized. 

In order to relate the results of the decay spawning dynamics to the experimental observables, the properties from the AIMD simulations are coupled to the above EOMs through the decay spawning algorithm described in the next section. To compute the population weight averaged properties that describe the overall dynamical behavior of the decay cascade wavepacket, we use the following equation, 
\begin{equation}\label{eq:charge_population}
\chi^{I,J}(t) =  \frac{\sum_{i}^{N(t)_{I}}  \alpha^{I,J}_{i}(t)p_{i}^{I,J}(t)    }{P^{I,J}(t)},
\end{equation}
where $\alpha^{I,J}_{i}$ represents any property (atomic charge, kinetic energy, bond length) of the decay spawning trajectory. $P^{I,J}(t)$ is the total state population and $\chi^{I,J}(t)$ is the population-weighted averaged property that will relate to the experimental observable.



\subsection{Computational Details of AIMD Simulations}\label{sec:comp}

\begin{table}[h]
  \caption{Table of active space, molecular symmetry group, spin state, number of averaged states, the state used in the gradient calculation (Rlx.) and the hole containing valence orbitals (V) used for each state in the model cascade.}
  \label{tbl:activespace}
  \begin{tabular}{@{\extracolsep{\fill}}lllllllll}
    \hline
      State & Active space & Sym. & Spin & State Avg. & Rlx. & V \\
    \hline
    (0,0): IBr[$3p_{Br}^{1}$]$^{+1}$         & RAS(15,1,1;1,7,0) & $a1$ & 2 & 1 & n/a & n/a\\
                                  & RAS1: 13$a1$ (HEXS) &  &  &  &\\
                                  & RAS2: 19$a1$, 20$a1$, 21$a1$, &  &  &  & & \\
                                  & 9$b1$, 10$b1$, 9$b2$, 10$b2$ &  &  &  & & \\
    (1,1): IBr[$3d_{Br}^{1}$,$\sigma_{pz}^{1}$]$^{+2}$  & RAS(14,1,1;1,7,0) & $a2$ & 3 & 3 & 1 & 21$a1$\\
                                  & RAS1: 2$a2$ (HEXS) &  &  &  & & \\  
                                  & RAS2: 19$a1$, 20$a1$, 21$a1$, &  &  & & & \\
                                  & 9$b1$, 10$b1$, 9$b2$, 10$b2$ &  &  & & & \\
    (1,2): IBr[$3d_{Br}^{1}$,$\sigma_{s}^{1}$]$^{+2}$ & RAS(12,2,1;2,5,0) & $a1$ & 3 & 1 & n/a & n/a\\
                                  & RAS1: 16$a1$, 19$a1$ (DEXS) &  &  & & & \\ 
                                  & RAS2: 21$a1$, 9$b1$, 10$b1$, &  &  & & & \\
                                  & 9$b2$, 10$b2$ &  &  &  & & \\
    (2,1): IBr[$\sigma_{s}^{1}$,$\sigma_{pz}^{1}$,$\pi_{px}^{1}$]$^{+3}$  & RAS(11,1,1;1,6,0) & $b1$ & 4 & 2 & 2 & 21$a1$, 9$b1$\\
                                  & RAS1: 19$a1$ (HEXS) &  &  &  & & \\  
                                  & RAS2: 20$a1$, 21$a1$, 9$b1$,&  &  &  & & \\
                                  & 10$b1$, 9$b2$, 10$b2$ &  &  &  & & \\
    (2,2): IBr[$\sigma_{s}^{2}$,$\sigma_{pz}^{1}$]$^{+3}$  & RAS(11,2,1;1,6,0) & $a1$ & 2 & 2 & 1 & 21$a1$\\
                                  & RAS1: 19$a1$ (DEXS) &  &  &  & & \\
                                  & RAS2: 20$a1$, 21$a1$, 9$b1$,&  &  &  & & \\
                                  & 10$b1$, 9$b2$, 10$b2$ &  &  &  & & \\
    \hline
  \end{tabular}
\end{table}

AIMD simulations for core-excited state dynamics in gas-phase water have been performed using density functional theory \cite{doi:10.1021/acs.jctc.8b00211}.  In this work, the electronic structures of all IBr potential energy surfaces are calculated by the restricted active space self-consistent field (RASSCF) method\cite{werner1981quadratically,malmqvist1990restricted} with second-order perturbation theory (RASPT2) to include dynamic correlation corrections to the energies\cite{malmqvist2008restricted} in the OpenMolcas electronic structure code\cite{fdez2019openmolcas}. In Table \ref{tbl:activespace}, we describe the active spaces used for each state involved in the model decay cascade, using the notation RASSCF($n$,$l$,$m$;$i$,$j$,$k$), where $n$, $l$, and $m$ refer to the total number of active electrons, the maximum number of holes allowed in RAS1, and the maximum number of electrons allowed in RAS3, respectively; $i$, $j$, and $k$ refer to the number of orbitals in the RAS1, RAS2, and RAS3 subspaces. The C2v point symmetry group was used in all cases. This restriction facilitated the production of smooth potential energy surfaces by limiting the configuration space during the AIMD simulation.  To select the hole location and generate the desired configuration, the target orbitals were placed into RAS1 with an individual sub-symmetry group applied to prevent orbital rotation during optimization. Then, the highly or doubly excited state (HEXS or DEXS) scheme \cite{delcey2019efficient} was applied to force the generation of either single or double hole occupations. The single-state implementation of RASPT2 (SS-RASPT2) was applied in all cases, using an imaginary shift of 0.1 a.u. Unless stated otherwise, all electronic structure calculations use the ANO-RCC-VQZP basis set. The use of this basis set in OpenMolcas includes scalar relativistic effects through atomic mean field integrals and the Douglas-Kroll-Hess (DKH) Hamiltonian. The 5th order DKH Hamiltonian with an exponential parameterization was used in this work. The atomic charges presented in this work are generated from the natural orbitals by the libwfa wavefunction analysis library \cite{https://doi.org/10.1002/wcms.1595} in OpenMolcas.

All AIMD simulations were initiated from the experimental ground state IBr structure with a bond length of 2.469 \AA. We excluded the temperature effect and set the initial velocities to zero. A time-step of 0.242 fs was used in the velocity Verlet algorithm, with velocities scaled to maintain constant total energy throughout the AIMD simulations. 
 Eqns. \ref{eq:eom1} and \ref{eq:eom2} were solved numerically using the fourth-order Runge–Kutta algorithm (RK4). The RK4 method can suffer from numerical population loss for large time-steps. Therefore, to minimize numerical error in the RK4 procedure, the AIMD potential energy curves were fitted with a 1D spline of 100,000 points.  The fitted potential energy surfaces were then used in the RK4 simulation with a timestep of 0.00242 fs and the time step and Auger-Meitner decay rates were scaled by 4 orders of magnitude during the RK4 procedure to further reduce the numerical error. 

\subsection{Decay Spawning Algorithm}\label{sec:algo}


\begin{figure*}[ht!]
 \centering
 \includegraphics[width=16cm]{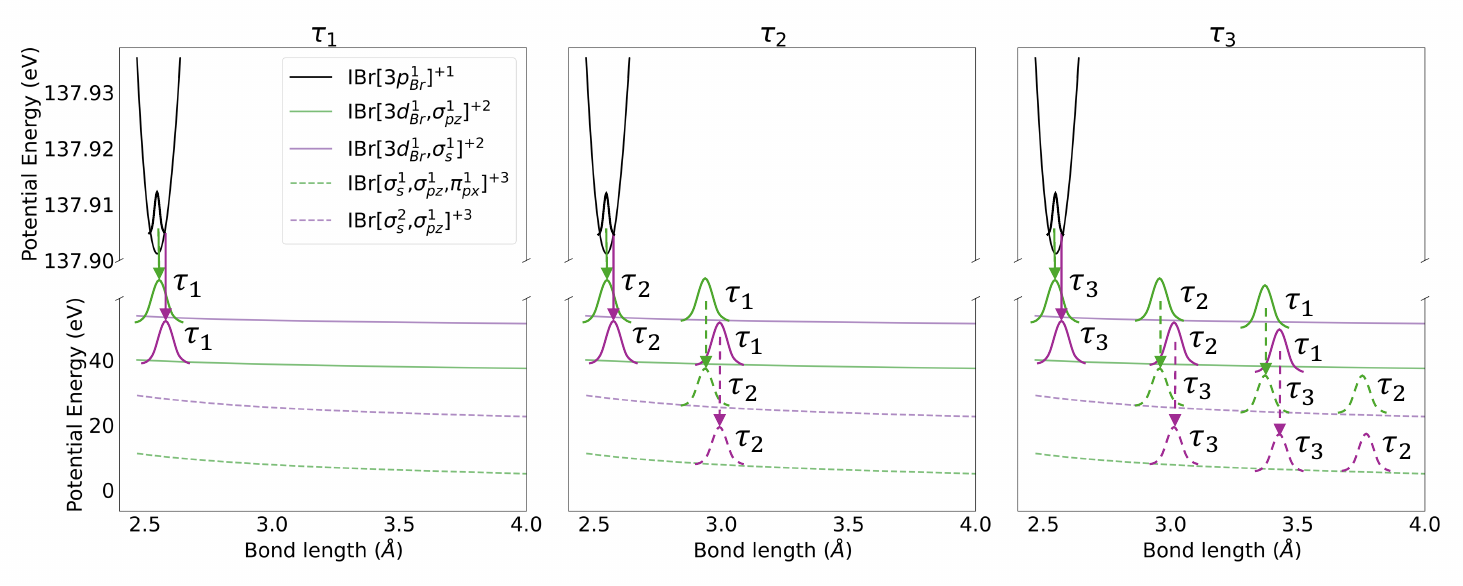}
 \caption{Schematic describing how the population is transferred across potential energy surfaces of increasing charge state via trajectory spawning for the first 3 spawning events, indicated by $\tau_{n}$.}
 \label{fgr:spawn_scheme}
\end{figure*}

Fig. \ref{fgr:spawn_scheme} depicts how the decay spawning algorithm transfers the population through the model decay cascade at the first three spawning events ($\tau_{1}$, $\tau_{1}$, $\tau_{3}$). All later spawning events will repeat $\tau_{3}$ with an expanding basis of quantum amplitude trajectories. 


\begin{figure*}[ht!]
 \centering
 \includegraphics[width=16cm]{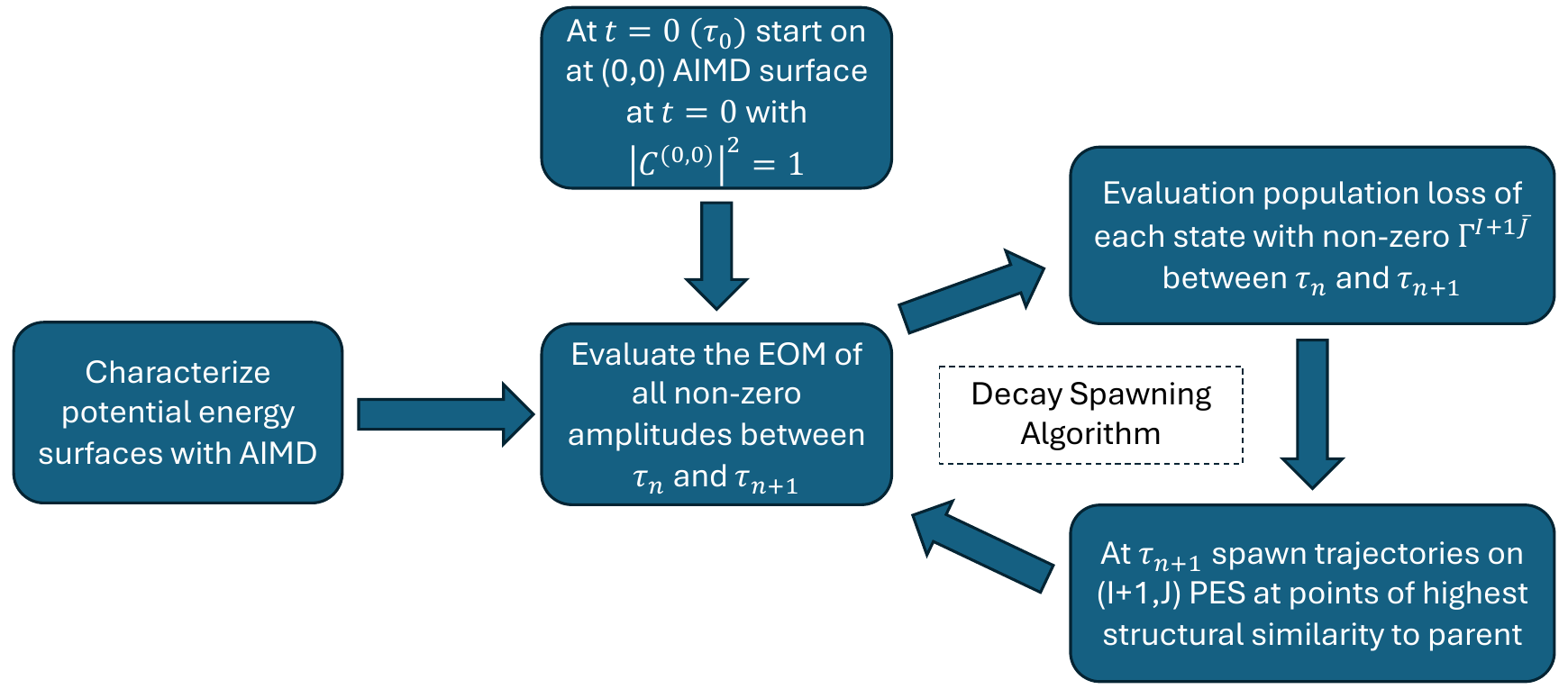}
 \caption{Flow diagram describing how the decay spawning steps are performed on potential energy surfaces pre-calculated by AIMD in this work.}
 \label{fgr:flow_diagram}
\end{figure*}

Once the potential energy surfaces have been characterized by AIMD, we initiate the spawning dynamics across the surfaces with the initial condition of $|C^{0,0}|^{2} = 1$ at $t=0$, where the spawning decay dynamics begins at $t=0$ of the AIMD simulation.  The flow diagram in Fig. \ref{fgr:flow_diagram} shows how the decay spawning algorithm can be executed in three steps.  These steps are applicable to any spawning event ($\tau_{n}$) for any number of decay cascade steps and channels:
\begin{enumerate}
  \item Evaluate the EOM of all non-zero amplitudes between $\tau_{n}$ and $\tau_{n+1}$. 
  \item Distribute the total population loss of each state with a non zero $\Gamma^{I+1,\bar{J}}$ into the population gains of the corresponding Z+1 states of each decay channel by the contributing partial decay rates ($\Gamma^{I+1,J}$) in the population loss function.
  \item At $\tau_{n+1}$, each (parent) trajectory (with $\Gamma^{I+1,\bar{J}} \neq 0$ ) that existed before $\tau_{n+1}$ spawns new (child) amplitude trajectories on a set of ($I+1$,$J$) potential energy surfaces.  
  The child amplitudes are the square root of the population loss values for each channel and the phase is inherited from the parent trajectory.
\end{enumerate}
Steps 1-3 can be repeated until the desired properties (state population, kinetic energy, atomic charge) are converged.  Since our current implementation of this method uses pre-calculated, discretized potential energy surfaces, the initial structures of the individual spawning child trajectories in Step 3 are chosen to have the highest structural similarity to the structure of the parent trajectory at the time of the spawning event. 
In the case of diatomic IBr, the structural similarity is simply defined as the closest bond length. For polyatomic systems, this metric could be defined by the root mean square deviation (RMSD). 

\section{Results and Discussion}\label{sec:results}

\subsection{AIMD}\label{sec:aimd}

\begin{figure}[ht!]
 \centering
 \includegraphics[width=15cm]{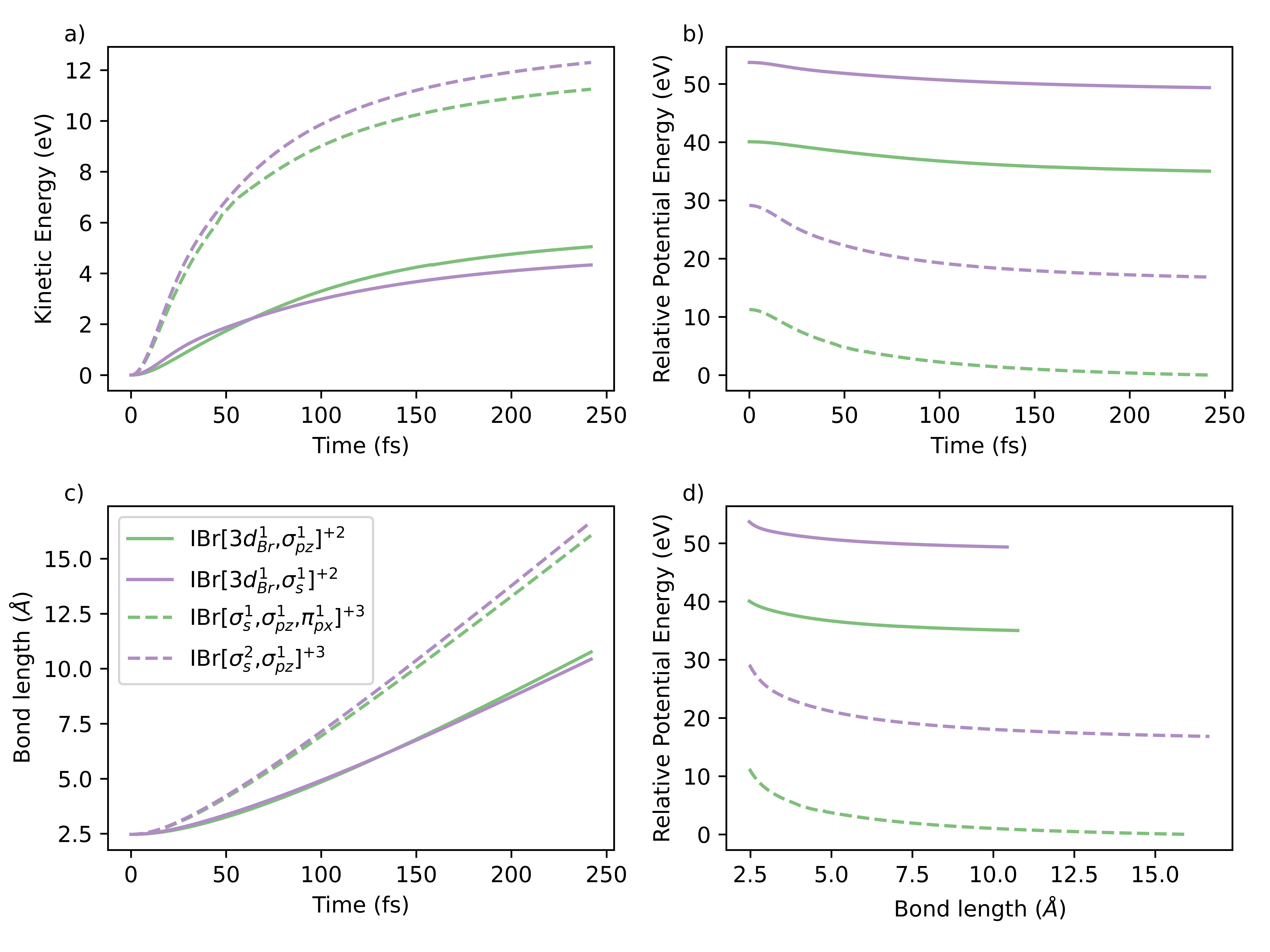}
 \caption{Energetic RASSCF/PT2 AIMD results for the dication (solid lines) and trication (dashed lines) states involved in the decay cascade model. Calculations used the ANO-RCC-VQZP basis set. The RASSCF active spaces are described in Table \ref{tbl:activespace}. No temperature effects were included, additional details are given in the Computational Details section.}
 \label{fgr:pes_aimd}
\end{figure}

We first discuss the AIMD results 
associated with the dissociation dynamics of the cations involved in the decay cascade. Fig. \ref{fgr:pes_aimd} a) shows the kinetic energy as a function of time.  The dication (solid lines) and trication (dashed lines) states converge towards kinetic energies of approximately 3-4 eV and 10-12 eV, respectively, with minor differences between the two channels.
Fig. \ref{fgr:pes_aimd} b) and c) show the relative potential energy and bond lengths over time. Since the simulations of these cation states are initiated at the ground state geometry with zero velocities, their initial kinetic energies and bond lengths are the same at $t = 0$.  The results indicate that changes in the kinetic energies and bond lengths are more affected by the charge state than by the selected electronic configurations within the charge states. Although there is a strong dependence on the initial potential energy with respect to both the charge state and electronic configuration, the change in potential energy reflects the kinetic energy as the velocity scaling maintains a constant total energy. Fig. \ref{fgr:pes_aimd} d) shows the relative potential energy surfaces of the cation states involved in the decay cascade as a function of bond length. We note that at the final simulation time of 242 fs the potential and kinetic energies have not fully converged as there remains a minor Coulombic interaction between the atomic fragments at the final bond distances.

\begin{figure}[ht!]
 \centering
 \includegraphics[width=15cm]{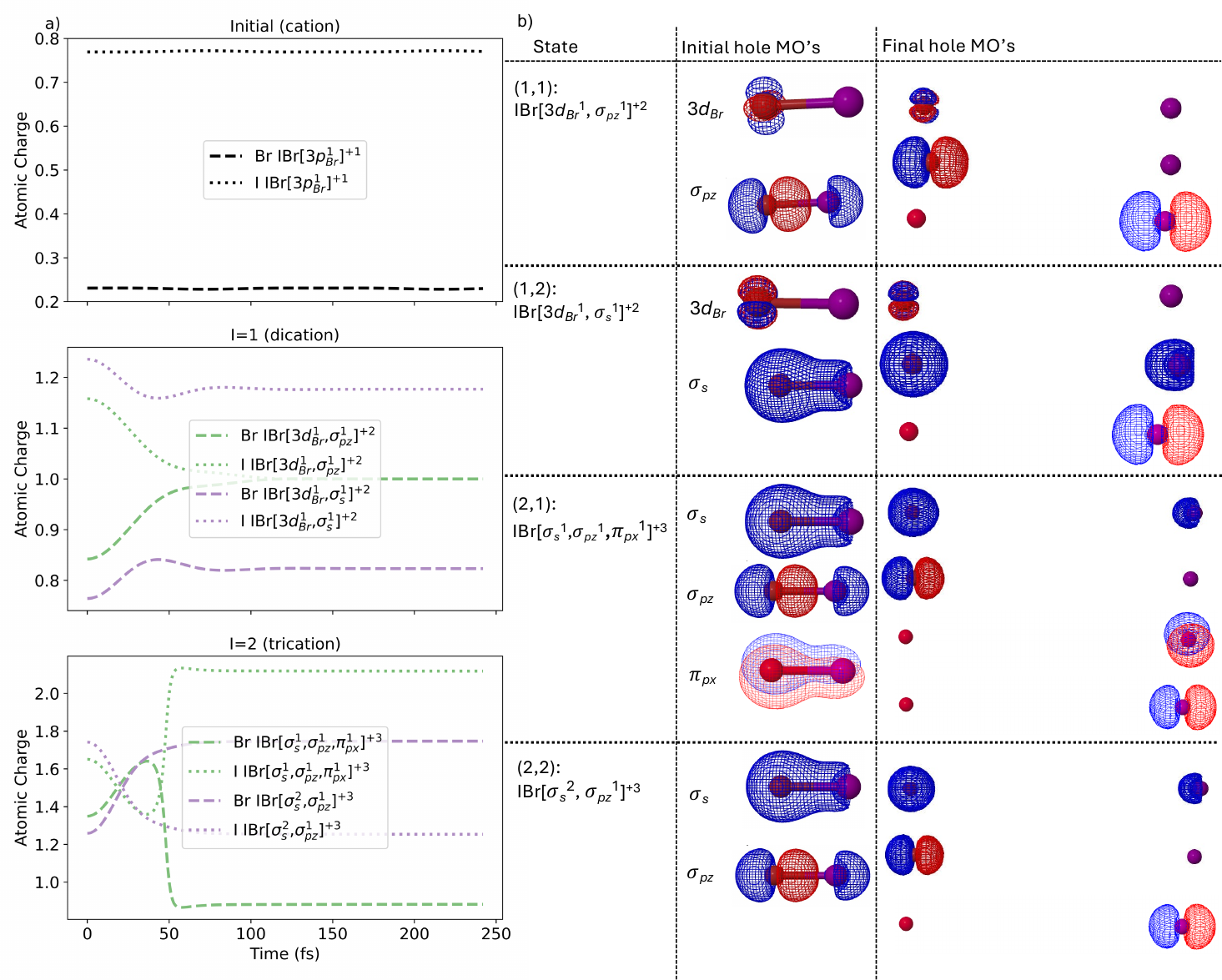}
 \caption{a) L\"{o}wdin charge  RASSCF/PT2 AIMD results for the Br (dashed lines) and I (dotted lines) atoms for all 5 states involved in the decay cascade model. Calculations used the ANO-RCC-VQZP basis set. b) shows the molecular hole orbitals in the initial and final geometries from the RASSCF calculations used in the AIMD simulations. The RASSCF active spaces are described in Table \ref{tbl:activespace}. No temperature effects were included, additional details are given in the Computational Details section.}
 \label{fgr:charge_aimd}
\end{figure}

Using L\"{o}wdin electron population analysis, Fig. \ref{fgr:charge_aimd} a) shows the atomic charge populations of the initial cation, dication and trication states. 
The dashed and dotted lines indicate the atomic populations on Br and I, respectively. 
In the initial state, the hole is located on the Br 3$p$. However, due to molecular bonding, the atomic charges for Br and I are approximately 0.2 and 0.8, respectively. 

For IBr[$3d_{Br}^{1}$,$\sigma_{pz}^{1}$]$^{+2}$ dication state, the atomic Br and I atomic charge converge to a 1 and 1 respectively.  On the other hand, for  IBr[$3d_{Br}^{1}$,$\sigma_{s}^{1}$]$^{+2}$ dication state, the atomic Br and I charge converge to 0.8 and 1.2 respectively. 
Similar fractional charges are shown in the dynamics of the trication states.  
For IBr[$\sigma_{s}^{1}$,$\sigma_{pz}^{1}$,$\pi_{px}^{1}$]$^{+3}$, it shows a charge oscillation between the Br and I sites during the first 50 fs before reaching a final atomic Br and I charge state of 2.1 and 0.9.  Also, for IBr[$\sigma_{s}^{2}$,$\sigma_{pz}^{1}$]$^{+3}$, it converges to the atomic Br and I charge  of 1.7 and 1.3 respectively.

We interpret the observed non-integer charges on the dissociated atoms as an effect of molecular bonding. Fig. \ref{fgr:charge_aimd}c shows the RASSCF molecular hole orbitals in the dication and trication states at both the initial and final structures of the AIMD simulations. It can be seen that the fragmentation of the $\sigma_{s}$ orbital results in hole density on both the Br and I atoms. This molecular orbital is formed from the Br 4\textit{s} and I 5\textit{s} orbitals, which therefore contribute to the molecular bonding at the level of theory used in this study. However, the non-integer charges can possibly be attributed to the finite set of orbitals used in the active spaces, the molecular symmetry restrictions adopted in our simplified model and the neglect of other decay processes, including fluorescence. Nonetheless, it is reasonable to interpret that the two dication states produce $I^+/Br^+$ ion pairs, whereas IBr[$\sigma_{s}^{2}$,$\sigma_{pz}^{1}$]$^{+3}$ and  IBr[$\sigma_{s}^{1}$,$\sigma_{pz}^{1}$,$\pi_{px}^{1}$]$^{+3}$ produce $I^+/Br^{2+}$ and $I^{2+}/Br^+$ ion pairs, respectively.
In the following section we show how these atomic charge populations can be coupled with the electronic state populations from the decay spawning dynamics to produce the overall atomic charge ratio which can be compared to experiment.

Before applying the AIMD fragmentation results to the decay spawning dynamics we briefly discuss the basis set and temperature effects on the AIMD results which are presented in the Appendix. In Fig. \ref{fgr:basis_set}, we show the effect of the basis set sizes, by comparing the double (VDZP), triple (VTZP) and quadruple (VQZP) zeta variations of the ANO-RCC basis set. The results show that all basis set sizes converge to the same charge distributions, therefore the final results of the spawning dynamics simulation will not be dependant on the basis set. This result also justifies using the VTZP basis set in the temperature effects simulations shown in Fig. \ref{fgr:temperature}. The figure presents six trajectories with initial velocities sampled from the Wigner distribution of the ground-state optimized geometry. 
The minimal variation between the trajectories demonstrates the suitability of using a single AIMD trajectory for each state in the decay spawning dynamics discussed in the next section.

\subsection{Decay Spawning Dynamics}\label{sec:spawn}


\begin{figure}[ht!]
 \centering
 \includegraphics[width=15cm]{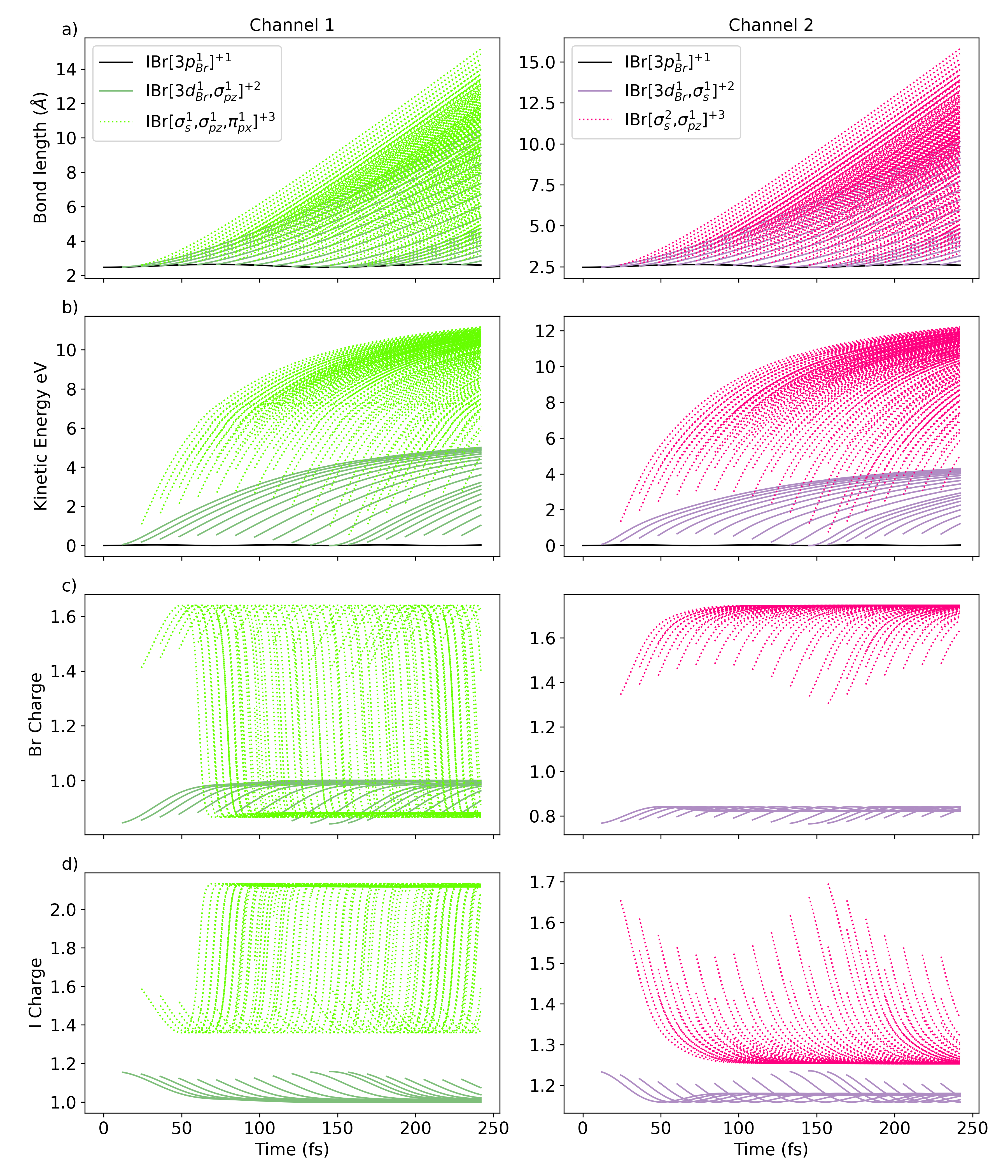}
 \caption{Evolution of the a) bondlength, b) kinetic energy, c) Br charge and d) I charge as a result of the decay spawning algorithm developed by this work. Simulations use the AIMD results with the ANO-VQZP basis without temperature effects presented in Fig.'s \ref{fgr:pes_aimd} and \ref{fgr:charge_aimd}.}
 \label{fgr:spawn_data}
\end{figure}

Fig. \ref{fgr:spawn_data} demonstrates dynamics of a sequence of 20 successive spawning events in both Channel 1 and 2, resulting in 20 trajectories in the dication states and 191 trajectories in the trication states in each channel.  
Fig. \ref{fgr:spawn_data} a) shows the change in bond lengths as new trajectories are spawned. The solid black line is the initial IBr[$3p_{Br}^{1}$]$^{+1}$ state.  At each spawning event, a single new trajectory is spawned on the dication AIMD surface (dark green solid lines for Channel 1, dark purple solid lines for Channel 2) at the point of closest bond length to the IBr[$3p_{Br}^{1}$]$^{+1}$ state at $\tau_{n}$. The spawning of trajectories determined by maximum structural similarity is then mapped to other molecular properties from the AIMD simulation. Fig. \ref{fgr:spawn_data} b) shows the kinetic energies of the spawning decay trajectories. Unlike the bond lengths, there is now a discontinuous evolution in the kinetic energy between different surfaces as a result of the trajectories spawning on potential energy surfaces with steeper energy gradients. Fig. \ref{fgr:spawn_data} c) and d) show the Br and I atomic charges respectively, they also have a discontinuous evolution as the total charge increases by 1 between the spawning events. 

\begin{figure}[ht!]
 \centering
 \includegraphics[width=15cm]{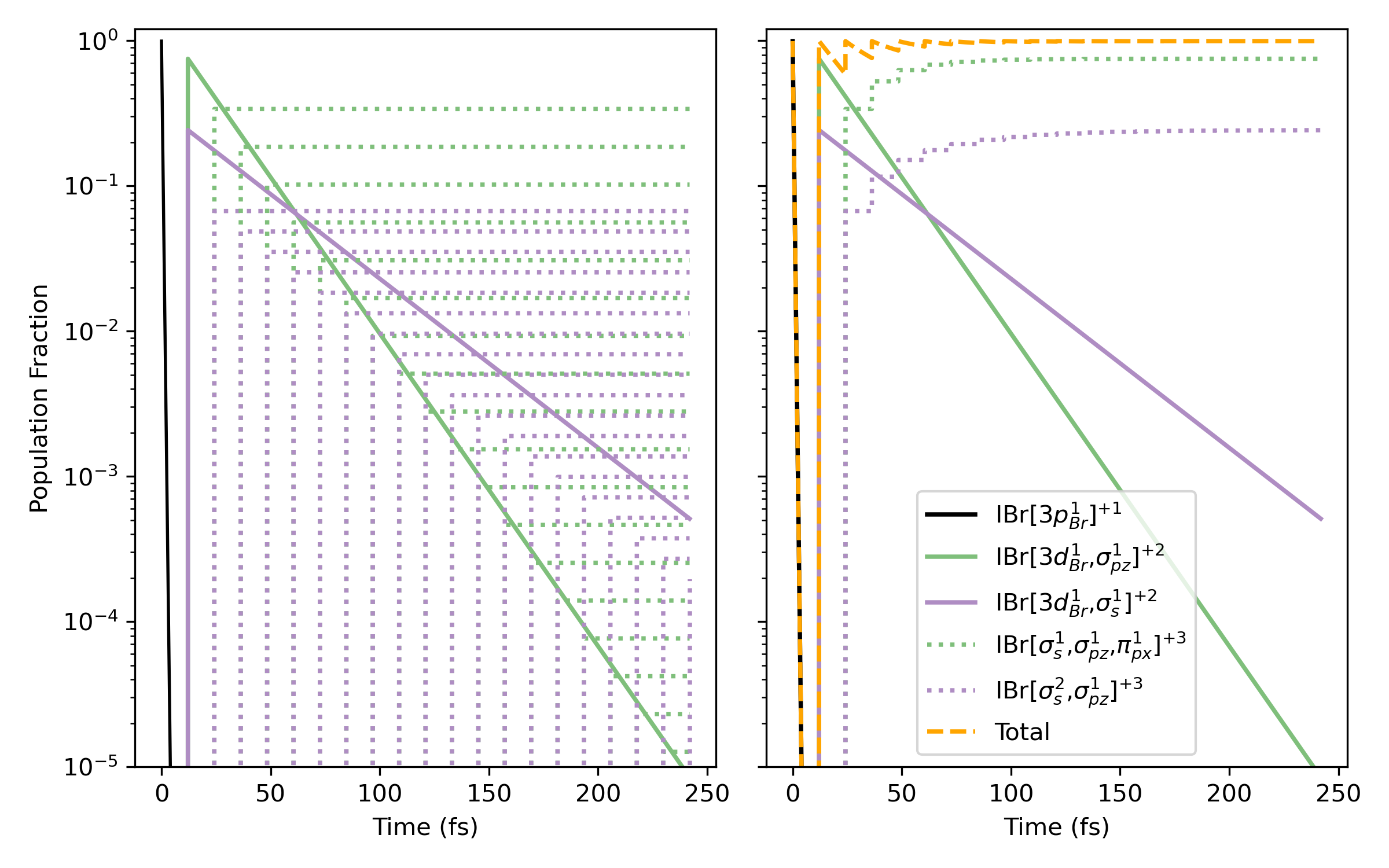}
 \caption{Time evolution of the state populations in the decay spawning dynamics. a) Shows the populations of each amplitude in the time expanding basis of trajectories. b) Sum of the trajectories for each state to give the total state population.}
 \label{fgr:state_population}
\end{figure}

To give the distribution of trajectories shown in Fig. \ref{fgr:spawn_data} physical meaning, we assign population weights from the expanding basis of quantum amplitudes trajectories from the spawning decay dynamics simulation. Fig. \ref{fgr:state_population} a) and b) show the evolution of the state populations over time for the individual amplitude trajectories and for the sum of these trajectories respectively. Between $t = 0$ and the first spawning decay event ($\tau_{1}$), the population of the initial IBr[$3p_{Br}^{1}$]$^{+1}$ state is depleted by the Auger-Meitner decay process to the two dictation states. At $\tau_{1}$, the population loss from the IBr[$3p_{Br}^{1}$]$^{+1}$ state is transferred to the dictation states, renormalizing the total population of the wavepacket to  1. Between $\tau_{1}$ and $\tau_{2}$, the population of the dication states is reduced, and at $\tau_{2}$, this population loss is transferred to the trication states. The population transfer from dication to trication states continues for the remainder of the simulation. 

From the ratio of the population of IBr[$\sigma_{s}^{1}$,$\sigma_{pz}^{1}$,$\pi_{px}^{1}$]$^{+3}$ to IBr[$\sigma_{s}^{2}$,$\sigma_{pz}^{1}$]$^{+3}$ at 242 fs shown in Fig. \ref{fgr:state_population}, our model shows branching ratio of $I^{2+}/Br^{+}$ and $I^{+}/Br^{2+}$ fragmentation channels to be about 3.  This calculated ratio is slightly higher than the experimental value of 1.5.  This overestimation can be due to our simplified model, which includes only 5 states.  

\begin{figure}[ht!]
 \centering
 \includegraphics[width=15cm]{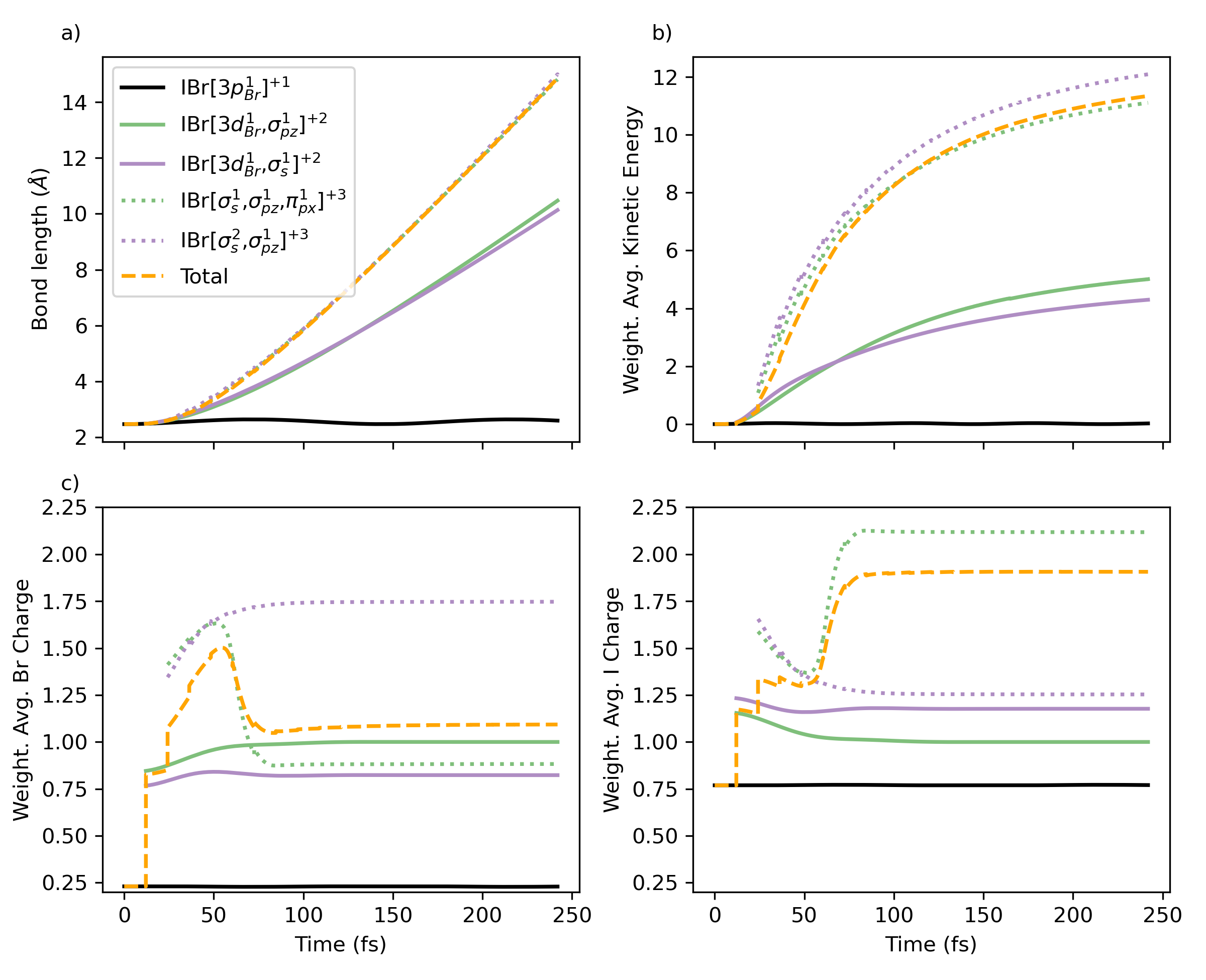}
 \caption{Population weight averaged properties of the cascade model dynamics produced by the decay spawning algorithm. Weight averaged a) bond-length b) kinetic energy and c) atomic charge results produced by applying the populations of the individual trajectories summed in a) to the results presented in Fig. \ref{fgr:spawn_data}.}
 \label{fgr:charge_pop}
\end{figure}

Fig. \ref{fgr:charge_pop} shows the population-weighted averages of bond lengths, kinetic energies, atomic charges, and spins derived from the results in Figs. \ref{fgr:spawn_data} and \ref{fgr:state_population}. It is clear the ion kinetic energies for both the dication and trication state do not yet converge at the end of our simulations (at 242 fs).  But, combining them with the bond length results, we deduce that the final total ion kinetic energies of $I^{2+}/Br^{+}$ and $I^{+}/Br^{2+}$ to be 14 and 13 eV respectively, which are in good agreement with the experimental values of 12 and 11 eV.


The evolution of the total population-weighted atomic charge in Fig. \ref{fgr:charge_pop} c) provides a quantum molecular charge transfer description of the multistep, multichannel Auger decay cascade.  Although a more frequent spawning recurrence would create a smoother charge transfer curve, the recurrence interval used in this simulation effectively captures the charge flow between the two atoms during the coupled Auger-Meitner decay and fragmentation dynamics. In the first 50 fs, the charge of IBr increases due to Auger-Meitner decays, followed by a charge flow towards the iodine atom. This process converges around 75 fs, which can be approximated as the total charge transfer time for this decay cascade model.


\section{Conclusion}

We have presented a novel approach for studying the multistep Auger-Meitner decay cascade dynamics following the core-ionization and fluorescence decay of heavy elements. Our approach combines AIMD with a multiple spawning method and provides a quantum molecular description of the decay process.  Previous MC/MD simulations using the COB model describe the electronic structure with the atomic HFS level of theory. Our current study includes the molecular bonding effects.
The COB model separates the charge transfer from the Auger-Meitner decay by describing charge transfer as an instantaneous process that fills the vacancy of a neighboring atom when the electron binding energy is higher than the Coulomb barrier. Our decay spawning dynamics approach goes beyond this by describing the charge transfer as a continuous process, modeling the gradual charge separation between the two atoms synchronously with the Auger-Meitner decay-induced charge increments. 

To demonstrate our the decay spawning algorithm and dynamics simulation, this study uses a simplified decay cascade model for IBr with a core hole in the Br 3p orbital and focuses on the subsequent fragmentation dynamics that generate ion pairs of $I^{+}/Br^{2+}$ and $I^{2+}/Br^{+}$. Our model predicts the ratios of these two ion pairs and their kinetic energy that agree with the experimental values.
Our current implementation uses a single non-interacting potential energy surface for one possible electronic configuration of each state and restricts the configuration space in the electronic structure calculation to ensure the production of smooth potential energy surfaces. In future implementations, the decay spawning algorithm can be coupled with ``on-the-fly" trajectory surface hopping\cite{10.1063/1.459170} or multiple cloning (MC) methods\cite{10.1063/1.4891530}.  Multiple cloning dynamics methods combine the time expanding basis in AIMS with the propagation of an average trajectory used in multiconfigurational Ehrenfest dynamics\cite{10.1063/1.4734313} and have been recently developed to treat a dense manifold of electronic states\cite{doi:10.1021/acs.jpclett.9b01902}. This approach could potentially enable modeling both the Auger-Meitner decay across multiple charge states and the non-adiabatic dynamics occurring between potential energy surfaces of the same charge state. Furthermore, by including more orbitals and decay steps, our approach would provide a powerful exploratory description of the fragmentation dynamics following inner-shell decay processes in heavy element-containing molecules.

\section*{acknowledgement}
This material is based on work supported by the U.S. Department of Energy, Office of Basic Energy Sciences, Division of Chemical Sciences, Geosciences, and Biosciences through Argonne National Laboratory. Argonne is a U.S. Department of Energy laboratory managed by UChicago Argonne, LLC, under contract DE-AC02-06CH11357. A. E. A. F. is grateful for the support from the Eric and Wendy Schmidt AI in Science Postdoctoral Fellowship, a Schmidt Futures Program.


\section{Appendix}

\subsection{Basis Set and Temperature Effect}\label{sec:basistemp}

\begin{figure}[ht!]
 \centering
 \includegraphics[width=15cm]{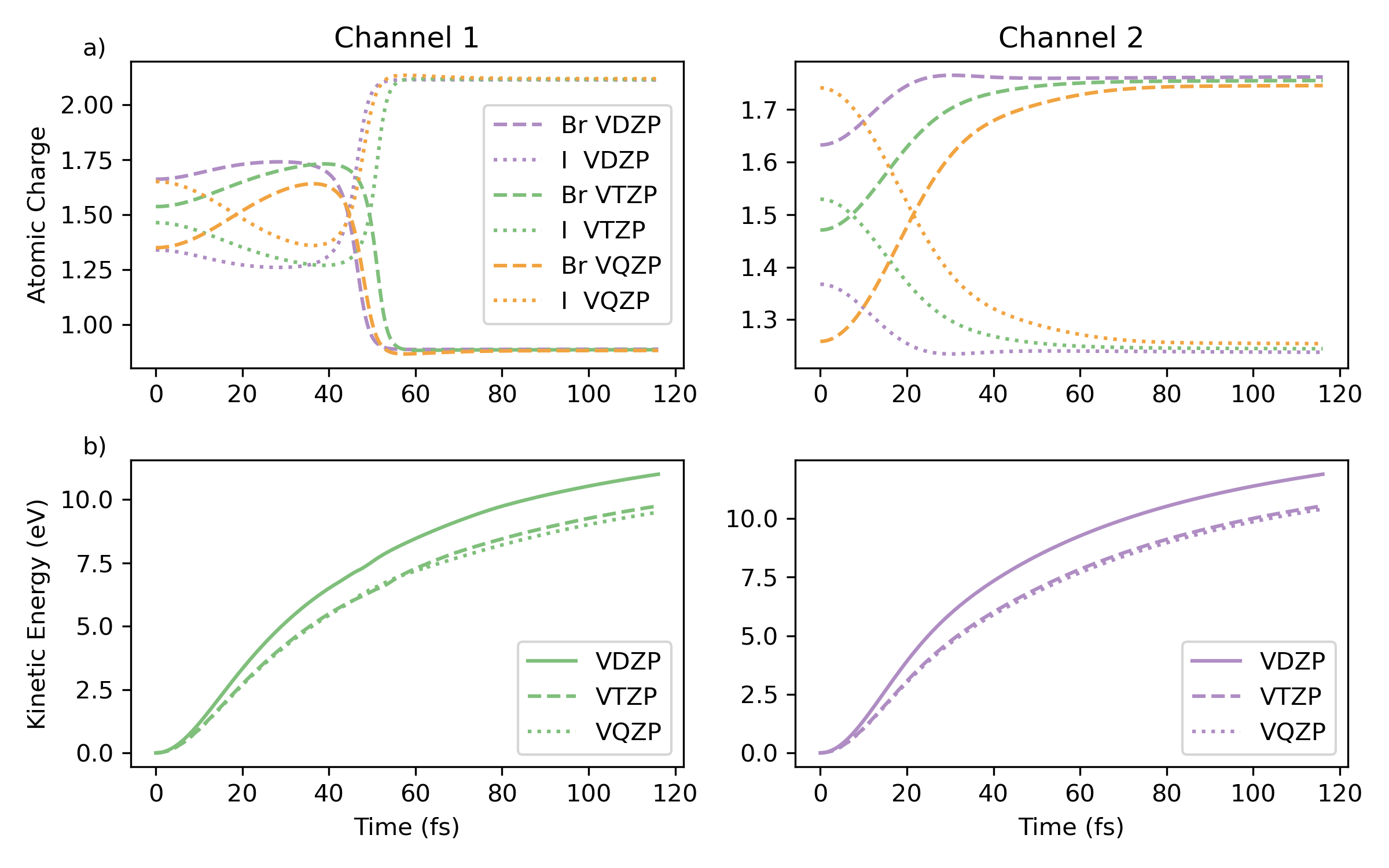}
 \caption{Effect of the basis set size on the RASSCF/PT2 AIMD results for the trication states involved in the decay cascade model. No temperature effects were included, additional details are given in the Computational Details Section.}
 \label{fgr:basis_set}
\end{figure}
Fig. \ref{fgr:basis_set} shows the dependence of the basis set size on the AIMD results. Panel (a) shows the atomic charge populations of the tri-cation states of Step 2 respectively. The results compare the double (VDZP), triple (VTZP) and quadruple (VQZP) zeta variations of the ANO-RCC basis set. The results demonstrate that all basis sizes converge to the same charge distributions. Therefore, the final results of the spawning dynamics simulation will not be dependant on the basis size in this case. However, discrepancies in the charge transfer dynamics below 50 fs in the simulation are evident, with a greater change in charge observed for larger basis sets. Fig. \ref{fgr:basis_set} c) shows the convergence of the kinetic energy with respect to the basis set size. Minimal difference is observed between the VTZP and VQZP basis sets, while the VDZP basis set yields a relatively higher final kinetic energy. This result justifies the use of the VTZP basis set, which has lower computational cost than the VQZP basis set.


\begin{figure}[ht!]
 \centering
 \includegraphics[width=15cm]{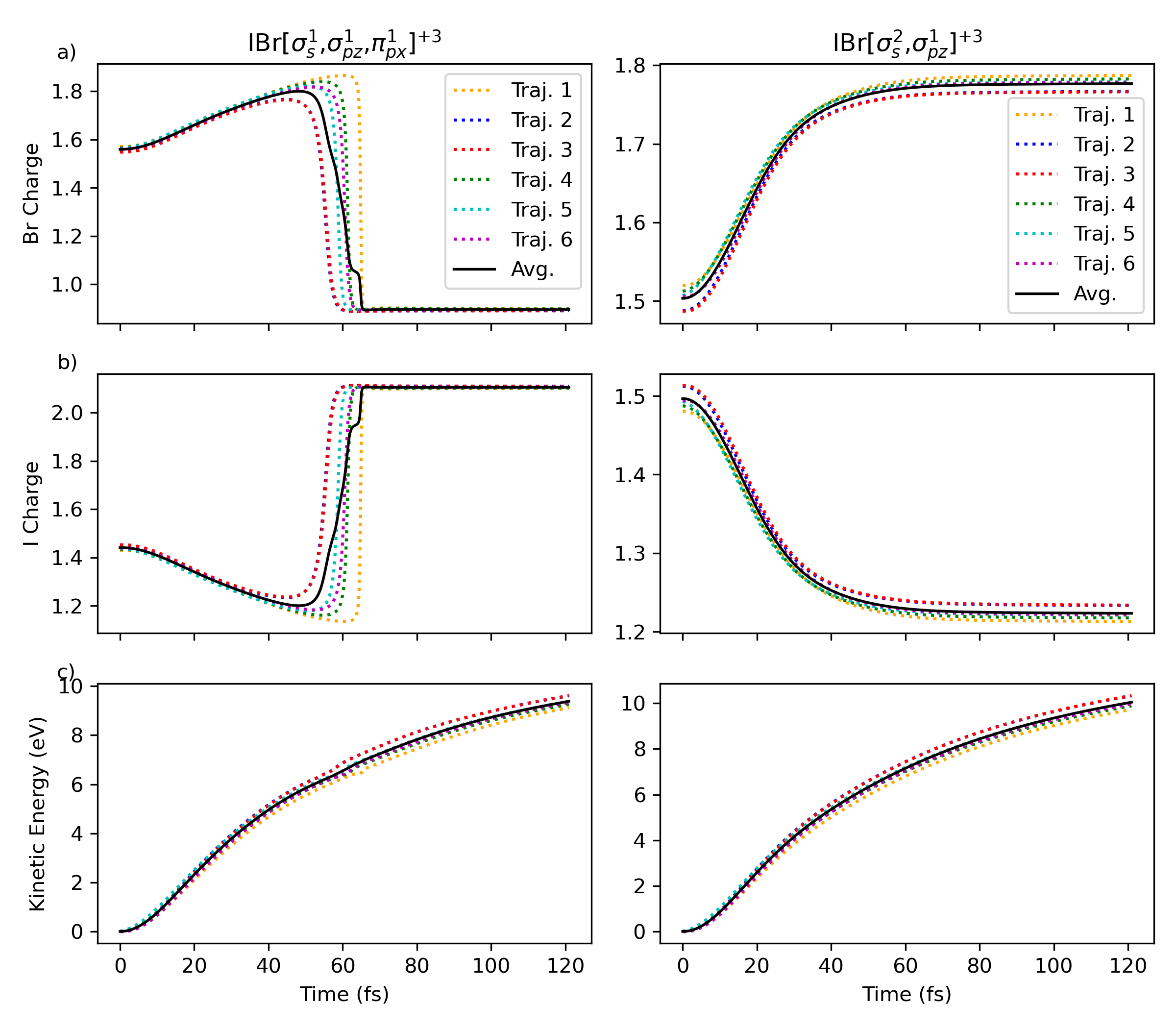}
 \caption{Temperature effecrs on the RASSCF/PT2 AIMD results for the dication and trication states involved in the decay cascade model. Calculations used the ANO-RCC-VTZP basis set. 6 trajectories using from initial geometries and velocities sampled from a Wigner distribution of the neutral IBr ground state optimised geometry were used. Additional details are given in the Computational Details Section. }
 \label{fgr:temperature}
\end{figure}

Using the VTZP basis set, we investigate the temperature effects on the AIMD results for the tri-cation states. Fig. \ref{fgr:temperature} displays six trajectories sampling initial velocities from the Wigner distribution of the ground state optimized geometry. It is evident that there is minimal variation among the trajectories at 300 K, and the final kinetic energy and charge separation results differ little from the VTZP results in Fig. \ref{fgr:basis_set}, which do not include any temperature effects and consider a single trajectory with an initial velocity of zero.






\bibliography{main}

\end{document}